\long\def\UN#1{$\underline{{\vphantom{\hbox{#1}}}\smash{\hbox{#1}}}$}
\def\NP{\vfil\eject}
\def\NI{\noindent}

\magnification=\magstep 1
\overfullrule=0pt
\hfuzz=16pt
\voffset=0.0 true in
\vsize=8.8 true in
\baselineskip 20pt
\parskip 6pt
\hoffset=0.1 true in
\hsize=6.3 true in
\nopagenumbers
\pageno=1
\footline={\hfil -- {\folio} -- \hfil}
\hphantom{A} 
\centerline{\UN{\bf Exactly Solvable Model of Quantum Spin}} 
\centerline{\UN{\bf Interacting with Spin Environment}}
\vskip 0.4in
\centerline{{\bf Dima Mozyrsky}}
\centerline{\sl Department of Physics, Clarkson University,}
\centerline{\sl Potsdam, NY 13699--5820}

\vskip 0.6in
 
\centerline{\bf ABSTRACT}

An exactly solvable model of a quantum spin interacting with a spin environment is considered. The interaction is chosen to be such that the state of the environment is conserved. The reduced density matrix of the spin is calculated for arbitrary coupling strength and arbitrary time. The stationary state of the spin is obtained explicitely 
in the $t \rightarrow \infty$ limit.

\NP

The problem of a quantum system interacting with a heat bath has been extensively studied
in various contexts during the last few decades. Originating from quantum optics in connection with studying spontanious emission and resonant adsorpsion [1-2], it has become an
important issue in condensed matter physics. The first, and probably one of the most famous
works in this field was by Caldeira and Leggett who studied effects of dissipation on the probability of quantum tunneling [3]. In this problem the heat bath is modeled by a set of
noninteracting harmonic oscillators linearly coupled to the quantum system. This model of heat bath has been mathematically justified in [4], and has become a widely accepted description of dissipative quantum dynamics, which advantageously combines both microscopical and phenomenological aspects of interaction between a quantum system and phonons or delocalized 
electrons [5]. A similar problem of a system in a double well potential under the
influence of a Heat Bath was studied in connection with magnetic flux tunneling in Josephson junctions [6]. It has been shown that the system looses its quantum coherence due to
the interaction with the heat bath and, in case of zero temperature and sufficiently strong coupling, completely localizes in one of the wells. Later this model was formalized by the spin-boson Hamiltonian [7,8] that received a lot of attention in modern condenced matter literature.
       
The study of quantum dissipation effects in the case of coupling to the 
fermionic heat bath has also received some attention in the literature [9]. For Hubbard-like coupling, behaviour quite different from that of the bosonic heat bath, has been found. The case of magnetic coupling
similar to RKKY interactions [10] was also extensively studied in conjecture 
with magnetic grains and giant spins of macromolecules interacting with a spin environment 
[11-13]. It is believed that this type of heat bath can not be mapped onto a bosonic bath model
and needs separate treatment [11,12]. This mechanism of quantum relaxation turns out to be 
effective especially at low temperatures resulting in a set of interesting phenomena, such as ``degeneracy blocking'' caused by the nuclear spins [11]. 

Another interesting and quite general effect that results in the ineraction of quantum system with its environment is  the destruction of quantum interference in the quantum system due to such interaction. This process, usually termed in the literature as decoherence, has attracted attention of both theorists and experimentalists not only due to its fundamental importance in quantum mechanics, but also due to the new fastly developing fields, such as quantum computing and quantum information theory, where decoherence is one of the major obstacles on the way of practical realizations of various, presently mostly hypothetical devices, such as quantum computers, etc [14]. No matter how well such a device is isolated from its environment, being essentially a macroscopic system, it will inevitably interact with the environment, resulting in the loss of interference between the states and thus desrupting its proper functioning. Several models have been proposed to study properties of decoherence in quantum computers and in more general systems [14-16]. The essential feature of these models is that the interaction
is set up in such a way that there is no energy exchange between the system under consideration and the environment, so the system's energy is conserved. This corresponds to a situation when a quantum system is very well isolated from its evironment, so only a phase exchange is allowed. One considers a hamiltonian 

$$ H = H_S + H_B + V \, , \eqno (1)$$

\NI 
where the first term $H_S$ in (1) corresponds to the system alone, $H_B$ is the heat bath and $V$ is the interaction between the bath and the system. It is assumed that the expectation value of $H_S$ is conserved during the system's evolution. This can be formalized by the assumption that the system's hamiltonian $H_S$ commutes with the full hamiltonian in (1), and in particular with the interaction term $V$. $H_S$ and $H_B$ naturally commute with each other as they act in different subspaces. This feature often allows one to carry out exact solutions for the system's reduced density matrix (to be defined later) in the basis of eigenavalues of the self-hamiltonain of the system $H_S$. 

In this work we consider an opposite extreeme, a model having a property that the state of the bath is preserved. The motivation to study such a model comes from a very common assumption in the literature on the propeties of the heat bath.        
It is often assumed that the heat bath has so many degrees of freedom that the effects of the interaction with the system dissipate away in it and will not influence the system back to any significant extent so that the bath remains described by a thermal equilibrium 
distribution at constant temperature, irrespective of the amount of energy and polarization
diffusing into it from the system [1,2]. This assumption is also called mollecular chaos or
Stosszahlansatz. There are some attempts to analyze this assumtion [17], but we beleive it is still little understood. This work is aiming to contribute to this topic.

Usually the following picture is assumed: the initial state of the full system is given by

$$ \rho (0) = \rho_S (0) \otimes \rho_B (0) \, , \eqno (2)$$

\NI
where $\rho_S(0)$ and $\rho_B(0)$ are the initial density matricies of the system and the bath 
respectively. The heat bath is initially assumed to be in thermal equilibrium state and the two systems are not initially entangled. When the interaction is switched on at time $t=0$, the full system's evolution is given approximately by the following 

$$ \rho (t) \simeq \rho_S (t) \otimes \rho_B (0) \, , \eqno (3)$$

\NI
that is, the state of the heat bath does not change in time to any significant extent.
This assertion, formalized by the Markoffian approximation, leads to the famous Pauli
master equations, which provide a key to understanding many profound phenomena in quantum
optics and condensed matter physics [1,2]. 

Here we propose a model for which equation [3]is exact.
Note that the form of interaction $V$ between the system and the bath has not been specified yet. Comparing equations (1) and (3), one can notice that in order for (3) to hold, one can simply require the following commutation relation

$$ [V\,,\,H_B] = [H\,,\,H_B] = 0 \, . \eqno (4) $$
   
Let us comment on this assumption. The most common form 
of the interaction between the system and the bath is the coupling between operators $Q_i$, acting in the subspace of the system and the bath operators $F_i$, so that

$$ V = \sum_i Q_i \otimes F_i \, . \eqno (5) $$

\NI
The choice of operators $Q_i$ and $F_i$ is usually determined by the particular features of 
the physical situation under consideration. However usually operators $Q_i$ and $F_i$ are not diagonal in the energy representation of the system and the bath and so the relation (4) may not be rigorously satisfied in general and is usually postulated by the Stosszahlansatz assumption.  
 
In this paper we study a model for which relation (4) is satisfied directly due to the commutation properties of the interaction and bath hamiltonians. Such a property, even though
not very common in the literature, seems worth exploring due to the above arguments. Moreover, this model allows {\it exact} solution for the reduced density matrix, which is a rather rare example in the literature on decoherence and thermalization of quantum systems.

In particular, we consider a model of interaction between a two-level quantum system and spin environment [11-14]. A bath of noninteracting spins $\vec \sigma^k$ is coupled to the 
two-level system $\vec \sigma^0$ under consideration. The coupling is chosen to be such that the Hamiltonian for the full system (two-level system + spin bath) is: 
 
$$ H=\Delta \sigma^0_z + \sum_k \omega_k\sigma^k_z + \sigma_x^0 \sum_k g_k\sigma_z^k \, , \eqno (6)$$

\NI
The first term in (6) corresponds to the two-level system and we will refer to it in the following as the central spin; the second is the spin bath, and the last one is the interaction between the bath and the two-level system. Here $2\Delta$ is the bare magnetic resonance (MR) frequency of the central spin, $\omega_k$ and $g_k$ are the frequencies and the coupling constants respectively for the spins of the spin bath. The spin bath self-hamiltonian obviosly commutes with the full hamiltonian in (6) and so the state of the bath is conserved in the course of the full system's evolution. 

The interaction is assumed to be switched on at time $t=0$ and the two systems (the central
spin and the bath) are initially not entangled with each other. The density matrix of the 
full system is given by

$$ \rho(0) = \left(|1_0 \rangle \langle 1_0|\right) \otimes {1 \over Z} e^{-\beta H_B} \, .\eqno (7) $$

\NI
Here $\beta = 1/k_BT$ is the inverse temperature and by $H_B$ we denote the self-hamiltonian of the bath, which is the second term in (6). The spin bath is assumed to be initially in thermal equilibrium at temperature $T$. $Z$ is the normalization constant or partition function of the free heat bath, i.e., a system of noninteracting spins

$$ Z = {\rm Tr} \, e^{-\beta H_B} = \prod_k \left(2 \cosh{\beta \omega_k}\right) \, . \eqno (8) $$

To avoid unnecesary mathematical complications we have assumed that the spin is initially in the excited state, even though the calculation can be carried out exactly for arbitrary
superposition of both ground and excited states.
The full system evolves in time $t$ quantum mechanically according to

$$ \rho (t) = U (t) \, \rho (0) \, U^{-1} (t) \, . \eqno (9) $$

\NI
Here and in the following we assume that $\hbar = 1$.

The evolution operator $U(t)=e^{-iHt}$ can be explicitely calculated by expanding the exponent
and combining the terms, that correspond to the expansion of cosine and sine 
respectively, resulting in

$$ U = e^{-H_B t} \left[ \cos{\gamma t} - i {{\Delta \sigma^0_z + \Omega \sigma^0_x}
\over \gamma} \sin{\gamma t} \right] \, , \eqno (10) $$

\NI
where
$$\Omega = \sum_k g_k \sigma^k_z \, , \eqno (11)$$
$$\gamma = \left( \Delta^2 + \Omega^2 \right)^{1\over 2} \, . \eqno (12)$$ 
\NI 
The reduced density matrix for the central spin is given by

$$ \rho^r (t) = {\rm Tr}^{\prime} \left[\rho (t)\right] \, , \eqno (13) $$
\NI
where prime denotes that the trace is taken over the states of the spin bath only.

We calculate the elements of the density matrix using the following technique. Consider the
magnetization of the central spin, which is given by the expectation value of $\sigma_z^0$
or the difference of the reduced density matrix diagonal elements

$$ \langle \sigma_z^0 (t) \rangle = \langle 1_0 | \rho^r (t) |1_0 \rangle -
\langle 0_0 |\rho^r (t)| 0_0 \rangle \, . \eqno (14) $$

\NI
With the use of (7), (9)-(10), equation (14), after some algebra, can be rewritten as

$$ \langle \sigma_z^0 (t) \rangle = {\rm Tr} \left[ {{e^{-\beta H_B}} \over Z}
\left( \cos^2{\gamma t} -{{\Omega^2 - \Delta^2} \over {\gamma^2}}\sin^2{\gamma t}
\right) \right] \, .\eqno (15) $$

\NI   
The trace in the basis of eigenstates of $H_B$ and $\Omega$ becomes a sum
over Ising-like variables $s_k=\pm 1$

$$ \langle \sigma_z^0 (t) \rangle = \sum_{\{s_k\}} {e^{-\beta H_B} \over Z}\,
\Lambda_{\eta} {{\sin{\gamma \eta}} \over {\gamma}}  \, , \eqno (16) $$

\NI
where the sum is taken over all possible configuartions of $s_k$. In equation (14) by $H_B$
and $\Omega$ (see eqs.(11)-(12)) we mean, of course, the eigenvalues of these operators, i.e., 
$H_B = \sum_k \omega_k s_k$ and $\Omega = \sum_k g_k s_k$. In (16) we have also intoduced 
operator $\Lambda_{\eta}$ given by

$$ \Lambda_{\eta} = \left[{{\partial} \over {\partial\eta}}\right]_{\eta = 2t} + 
\Delta^2 \int_0^{2t} d\eta  \, , \eqno (17) $$

\NI
which can be conviniently interchanged with the summation in (16).
In order to compute the sum (16) we employ the following identity [18]:

$$ {\rm Re} \int_0^{\eta}e^{ix\Omega}J_0\left(\Delta\sqrt{\eta^2 - x^2}\right) dx = 
{{\sin{\eta\left(\Delta^2 + \Omega^2\right)^{1\over2}}} \over {\left(\Delta^2 + \Omega^2\right)^{1\over2}}} \, . \eqno (18)$$

\NI
Here $J_0(z)$ is the zeroth order Bessel function. 
At this point performing the summation in (16) is a straightforward procedure as $\Omega$ enters linearly in the exponent in (18) and the sum is equivalent to the simple calculation of 
the partition function for a system of uncoupled spins in external magnetic field.
So using (18) and also (11)-(12), after some algebra we obtain:
 
$$ \langle \sigma_z^0 (t) \rangle = \Lambda_{\eta}\, {\rm Re} \int_0^{\eta} dx
J_0\left(\Delta\sqrt{\eta^2 - x^2}\right) {1 \over Z}\sum_{\{s_k\}}
e^{-\beta H_B + ix\Omega} =$$
$$ \Lambda_{\eta}\, {\rm Re} \int_0^{\eta} dx J_0\left(\Delta\sqrt{\eta^2 - x^2}\right)  
\Phi (x) \, ,\eqno (19) $$
and
$$ \Phi (x) = \prod_k \left[\cos{g_k x} - i\tanh{\beta\omega_k}\sin{g_k x}\right] \, .
\eqno (20) $$

\NI
The trick used in  (18)-(20) is simular to that in calculation of partition functions for certain mean field models, such as Curie-Weiss model; see [19] and references therein. 

The off-diagonal elements of the reduced density matrix can be calculated in a simular way. The final result is

$$ \rho^r_{10}(t) = \Lambda^{\prime}_{\eta} {\rm Im} \left[ \Phi (x) - \Delta \int_0^{\eta} x 
{{J_1 \left(\Delta \sqrt{\eta^2 - x^2}\right)} \over {\sqrt{\eta^2 - x^2}}} \Phi (x)\,dx 
\right] \, , \eqno (21) $$

\NI
where 

$$ \Lambda^{\prime}_{\eta} = \left[{i\over2}\right]_{\eta = 2t} + \Delta \int_0^{2t}
d\eta \, . \eqno (22) $$

\NI
Equations (19)-(21) constitute the main result of this work. All calculations
up to this point were {\it exact} for arbitrary coupling constants and energy splittings of the spins. 

Let us specify the coupling constants $g_k$ and MR frequencies $2\omega_k$ of the external spins
$\vec \sigma^k$. For simplicity we assume that the energy splittings for all
"external" spins are the same $\omega_k = \omega$ and the coupling constants are randomly 
distributed with average $\langle g \rangle$ and second moment $\langle g^2
\rangle$. Expression (20) for $\Phi (x)$ can be exponentiated thus transforming the
product into summation in the exponent
$$ \Phi (x) = \exp{\left[ \sum_k A_k (x)\right]} \, ,\eqno (23) $$
where
$$ A_k (x)={1\over 2}\ln{\left[\cos^2{g_k x} + \tanh^2{\beta\omega}\sin^2{g_k x}\right]}
- i\tan^{-1}{\left[\tanh{\beta\omega} \tan{g_k x}\right]} \, .\eqno (24)$$

\NI
Assume now that the sum in (23) contains $N$ terms, that is, the spin bath consists of $N$
spins, where $N \gg 1$. $A_k (x)$ are random numbers with some average $\langle A(x)
\rangle$. with this assumption expression (23) becomes

$$ \Phi (x) = \exp{N \langle A(x) \rangle} \, , \eqno (25) $$

\NI
where $\langle \ \rangle$ denote the average taken over the coupling constants.
In this work we assume that $ \langle g \rangle = 0$ and $ \langle g^2 \rangle = { C \over N}$,
where $C$ is of order unity. This choice is made only for simplicity of calculations and obviosly other possibilites for distributions of $\omega_k$ and $g_k$ can be explored.
 
Averaging of (24) can be done easily by expanding it up to the second order in $g$
with the result

$$ \Phi (x) = \exp{\left[-{C \over {2\cosh^2{\beta\omega}}} x^2 \right]} \, . \eqno (26) $$

\NI 
This relation, when inserted into (19)-(21), gives the explicit analytical expressions
for the density matrix of the spin. The magnetization (the difference between the 
diagonal elements) is an oscillatory function, which decays to the limiting value for
$\langle \sigma_z^0(\infty) \rangle$. This can be explicitly calculated and after
the straightforward manipulations such as change of order of integration in (19) one obtains:

$$ \langle\sigma^0_z(\infty)\rangle =\Delta \cosh{\beta\omega}\sqrt{{\pi\over 8 C}}\exp{\left(z^2\right)}\,erfc(z)  \eqno (27)$$

\NI
where

$$z = {\Delta \over \sqrt{2C}}\cosh{\beta\omega}  \, . \eqno (29)$$

\NI
Here $erfc(z)$ is the complimentary error function. A similar straightforward calculation shows that the off-diagonal elements vanish for $t=\infty$.

One could expect that the central spin ``thermalizes'' due to the interaction with the spin
bath, i.e., its density matrix reaches the state distributed according to the Boltzmann
distribution.  One can see that for this model this is not the case. Moreover, it is a straightforward observation that $\rho^r(\infty)$ depends on the initial state $\rho^r(0)$ and thus the conservation of the state of the bath is not sufficient to represent molecular chaos assumption and cannot represent a ``true'' heat bath. However its possible that by introducing a certain frequency dependence of the coupling constants $g_k$ one can obtain a reasonable approximation of a heat bath on a short time scale. Such  possibility should be subjected to further study.

In summary, we derived the exact results for the reduced density matrix of a spin
interacting with a type of spin bath. The precise functional dependence is determined
by the choice of the spin bath dispersion relation and its coupling to the ``central''
spin. It turns out that the system (spin) does not reach the canonical distribution in
the course of its evolution, and its stationary state for $t=\infty$ depends on the initial
conditions.

The author would like to thank Professors V. Privman and L.S. Schulman for their interest in the work and numerous fruitfull discussions. 

\NP

\centerline{\bf{References}}
{\frenchspacing

\item{[1]} K. Blum, {\it Density Matrix Theory and Applications}, Plenum Press, (1996).
 
\item{[2]} W.H. Louisell, {\it Quantum Statistical Properties of Radiation}, John Wiley \&
Sons, (1973). 

\item{[3]} A.O. Caldeira and A.J. Leggett,Phys. Rev. Lett. {\bf 46}, 211 (1981).

\item{[4]} G.W. Ford, M. Kac and P. Mazur, J. Math. Phys.{\bf 6}, 504 (1965).

\item{[5]} A.J. Legget in {\it Percolation, Localization and
Superconductivity}, 
NATO ASI Series B: Physics, Vol. {\bf 109}, edited by A.M. Goldman and S.A. Wolf 
(Plenum, New York 1984), p.1.

\item{[6]} A.J. Bray and M.A. Moore, Phys. Rev. Lett.
{\bf 49}, 1546 (1982).

\item{[7]} S. Chakravarty and A.J. Leggett, Phys. Rev. Lett.{\bf 52}, 5 (1984).

\item{[8]} Review: A.J. Legget, S. Chakravarty, A.T. Dorsey, M.P.A. Fisher and W. Zwerger, 
Rev. Mod. Phys. {\bf 59}, 1 (1987).

\item{[9]} L.-D. Chang, S. Chakravarty, Phys. Rev. B {\bf 31}, 154 (1985).

\item{[10]} J. Winter, {\it Magnetic Resonance in Metals}, Oxford at the Clarendon Press,
(1971).

\item{[11]} N.V. Prokof'ev and P.C.E. Stamp, J. Low Temp. Phys. {\bf 104}, 143 (1996).

\item{[12]} I.S. Tupitsyn, N.V. Prokof'ev, P.C. Stamp, {\it Effective Hamiltonian in the 
Problem of a ``Central Spin'' Coupled to Spin Environment}, (preprint).

\item{[13]} S. Sachdev and R.N. Bhatt, J. Appl. Phys. {\bf 61}, 4366 (1987).

\item{[14]} G.M. Palma, K.A. Suominen and A.K. Ekert, Proc. Royal. Soc. London A {\bf 452}, 567 (1996).

\item{[15]} W.G. Unruh, Phys. Rev. A {\bf 51}, 992 (1995).

\item{[16]} D. Mozyrsky and V. Privman, J. Stat. Phys. {\bf 91}, 787 (1998). 

\item{[17]} N.G. van Kampen, J. Stat. Phys. {\bf 78}, 299 (1995).

\item{[18]} I.S. Gradshtein, I.M. Ryzhik, {\it Table of Integrals, Series and Products},
Academic Press, Inc., (1980).

\item{[19]} C.J. Thompson, {\it Mathematical Statistical Mechanics}, The Macmillan Company, (1972).

}
\bye